\documentclass[a4paper]{jpconf}

\usepackage{mathrsfs}
\usepackage{cancel} 
\usepackage{subfigure}
\usepackage{amsmath,latexsym}

\begin{document}
\title{Thermodynamical aspects of black holes in modified gravity}

\author{Lorenzo Sebastiani}

\address{TIFPA - INFN,  Via Sommarive 14, 38123 Povo (TN), Italy\\
Dipartimento di Fisica, Universit\`a di Trento,Via Sommarive 14, 38123 Povo (TN), Italy
}

\ead{lorenzo.sebastiani@unitn.it}

\begin{abstract}
We discuss some thermodynamical definitions for black holes in modified theories of gravity.
\end{abstract}

\section{Introduction}
In General Relativity (GR), several thermodynamical notions can be introduced 
for the black holes (BHs), but in the modified theories of gravity the black hole solutions are not expected to share the same proprieties of their Einsteinian counterparts. 
In $F(R)$-modified gravity the First law of thermodynamics can be derived from the equations of motion, evaluating independently the entropy via Wald method and the Killing-Hawking 
temperature from the metric, and an expression for the BH Killing energy can be found.  
In an analogue way, in other theories of modified gravity (for instance, in Gauss-Bonnet gravity) the First Law of 
thermodynamics can be used to infer the black hole energy. 

This proceeding is mainly based on Refs.~\cite{{uno, due, tre}}.


\section{Black holes in General Relativity}

Any spherically symmetric and four dimensional metric can be locally expressed in the form:   
\begin{equation}
ds^2 =\gamma_{ij}(x^i)dx^idx^j+ {\mathcal R}^2(x^i) d\Omega_2^2\,,\qquad i,j \in \{0,1\}\,,
\end{equation}
where $d\Omega_2^2$ is the metric of a two-dimensional maximally symmetric space, $\gamma_{ij}(x^i)$ is the reduced metric of the normal space-time with coordinates $x^{i}$ and $\mathcal R(x^i)$
is the areal radius and is a function of the coordinates of the normal space. 
On the normal space one can 
introduce the scalar quantity 
\begin{equation}
\chi(x^i)=\gamma^{ij}(x^i)\partial_i  {\mathcal R}(x^i)\partial_j  {\mathcal R}(x^i)\,,
\end{equation} 
such that the sphere with areal radius $\mathcal R(x^i)$ turns out to be trapped when $\chi(x^i)<0$;
marginal when $\chi(x^i)=0$;
untrapped when  $0<\chi(x^i)$.
Thus, the dynamical trapping horizon of a black hole is defined by the conditions
\begin{equation} 
\chi(x^i)\Big\vert_H = 0\,,\quad0<\partial_i\chi(x^i)|_H\,.
\end{equation}
In this paper, the pedex `$H$' denotes a quantity evaluated on the coordinates of the horizon.

In General Relativity we can associate to the black hole horizons several thermodynamical quantities, namely the energy, the entropy and the surface gravity. For the energy
we have a quasi-local definition given by the Misner-Sharp formula,
\begin{equation}
E_{MS}(x^i):=\frac{1}{2G_N}{\mathcal R}(x^i)\left[1-\chi(x^i) \right]\,,
\end{equation}
with $G_N$ the Newton's constant.
Thus, the Misner-Sharp energy evaluated on the BH horizon corresponds to the BH Killing energy/mass,
\begin{equation}
E=\frac{r_H}{2G_N}\,.\label{BHEnergy} 
\end{equation}
The entropy of a black hole satisfies the Area Law,
\begin{equation}
S=\frac{\mathcal{A}_H}{(4 G_N)}\,,\label{AreaLaw}
\end{equation}
namely is proportional to the area $\mathcal A_H$ of the horizon.

Finally, for static black holes we may use 
the time-like Killing vector field $\xi_{\mu}(x^\nu)$ to define the Killing surface gravity $\kappa_K$ as follows,
\begin{equation}
\kappa_K\xi^{\mu}(x^\nu)=\xi^{\nu}\nabla_\nu\xi^\mu(x^\nu)\,.
\end{equation} 
In the dynamical case, where the time-like Killing vector field is absent, Hayward found a way to infer the surface gravity by working with the metric only,
\begin{equation}
\kappa_H:=\frac{1}{2}\Box_{\gamma} {\mathcal R(x^i)}\Big\vert_H\,, 
\end{equation}  
where the d'Alambertian is evaluated with respect to the reduced metric $\gamma_{ij}(x^i)$.

To the horizon of a black hole is also possible to associate a temperature. 
In fact the black holes are not so black and may emit radiation, dubbed the ``Hawking radiation'', due
to the quantum effects near to the horizon~\cite{HT}. 
In the static case, all derivations of the Hawking radiation rate
leads to the 
semi-classical expression,
\begin{equation}
\Gamma\equiv\mathrm{e}^{-\frac{2\pi\Delta E_K}{\kappa_K}}\,,
\label{ratekill}
\end{equation}
in terms of the change $\Delta E_K$ of the Killing energy of the emitted particle and the Killing surface gravity. Thus, the surface gravity can be identified with the Killing temperature as
\begin{equation}
T_K:=\frac{\kappa_K}{2\pi}\,.
\end{equation}
Therefore, if one uses the change of the entropy $\Delta S$ one easily obtains the First Law of black hole thermodynamics,
\begin{equation}
\Delta E_K=T_{K} \Delta S\,. 
\end{equation}
In the dynamical case Hayward found a way to derive the First Law from the equations of motion.
Assuming the Einstein's equation of GR, in a generic four-dimensional 
spherically symmetric space-time, the following geometric identity holds true on the black hole trapping horizon~\cite{Hay}, 
\begin{equation}
\frac{\kappa_{H}}{2\pi} \frac{d}{d r_H}\left(\frac{\mathcal A_H}{4G_N}\right)=\frac{d}{d r_H}\left(\frac{\mathcal R_H}{2G_N}\right)+
\frac{ \mathrm{T}_H^{(2)}}{2} \frac{d}{d r_H} \mathcal V_H\,, 
\end{equation}
where $\mathcal V_H$ is the three-volume enclosed by the horizon and 
$\mathrm{T}^{(2)}_H$ is the reduced trace of the matter stress-energy tensor at the horizon and acts like a working term.
On thermal equilibrium the Gibbs equation leads to,
\begin{equation}
T \Delta S =\Delta E
+ p dV\,,\quad
dV=\mathcal V_k r_H^2dr_H\,,
\end{equation}
such that,
by introducing  the entropy (\ref{AreaLaw}) and the BH energy (\ref{BHEnergy}), one may suggest the Kodama/Hayward temperature,
\begin{equation}
T_H:=\frac{\kappa_H}{2\pi}\,.
\end{equation}
Let us restrict our analysis to the spherically symmetric static space-time. The metric reads,
\begin{equation}
ds^2=-\mathrm{e}^{2\alpha(r)}B(r)dt^2+\frac{dr^2}{B(r)}+r^2d\Omega_k^2\,,
\quad d\Omega_k^2=\left(\frac{d\rho^2}{1-k\rho^2}+\rho^2 d\phi^2\right)\,,\label{metric}
\end{equation}
where $\alpha(r)$ and $B(r)$ are functions of the radial coordinate only, $\mathcal R=r$ is the areal radius and the topology depends on the $k$ parameter and can be 
spherical, flat or hyperbolic for $k=+1,0,-1$, respectively. A
static solution describes a black hole as soon as there exists an event horizon with a real and positive radius $r=r_H$ where
\begin{eqnarray}
B(r_H)=0\,,\quad
0<B'(r)|_{r_H}\,.
\end{eqnarray}
The prime index corresponds to the derivative with respect to $r$.
We should note that in the static case the Killing temperature $T_K$ and the Kodama temperature $T_H$ associated to the event horizon are in principle different when $\alpha(r_H)\neq 0$,
\begin{equation}
T_K:=\frac{1}{4\pi}\mathrm{e}^{\alpha(r_H)}B'(r_H)\,,
\quad T_H:=\frac{1}{4\pi}B'(r_H)\,.
\label{TK}
\end{equation}
In General Relativity this is not a problem. The Hawking radiation rate is independent on the choices of temperature and energy of the emitted particle and in the vacuum case 
of the Schwarzshild solution one has $\alpha(r)=0$ and the two definitions coincide.
However, this is not true for the vacuum case of a modified gravity theory where $\alpha(r)\neq 0$. Moreover, in modified gravity is not easy to define the energy of a black hole.
In General Relativity the Misner-Sharp mass corresponds to the charge of a conserved current from the second-order differential equations of the theory, but in modified gravity
we deal with higher derivative field equations and we must use a different approach. In what follows, we will consider some classes of modified theories
with black hole solutions and we will analyze 
the First Law of Thermodynamics in their framework.

\section{$F(R)$-four dimensional modified gravity}

Let us consider the $F(R)$-gravity  in vacuum, whose action is given by (see Ref.~\cite{Od} for some general reviews),
\begin{equation}
I=\frac{1}{16\pi G_N}\int_\mathcal{M} d^4 x\sqrt{-g} F(R)\,.
\end{equation}
Here, $g$ is the determinant of the metric tensor $g_{\mu\nu}(x^\mu)$, $F(R)$ is a function of the Ricci scalar only and $\mathcal M$ is the space-time manifold.
Given a static black hole solution described by the metric (\ref{metric}),
iff $R_H$ explicitly depends on $r_H$ only, from the (0,0)-component of the $F(R)$-field equations evaluated on the event horizon we obtain,
\begin{equation}
T_k\Delta S_W
=
\mathrm{e}^{\alpha(r_H)}\mathcal V_k\left(\frac{k\,F_R(R_H)}{2G_N}-\frac{R_HF_R(R_H)-F(R_H)}{4G_N}r_H^2\right)\,,
\label{EquazionePrincipe}
\end{equation}
where $\mathcal V_k\equiv \mathcal A_k/r^2$ depends on the topology, the Killing temperature $T_K$ (\ref{TK}) emerges in a natural way, and $S_W$ is the Wald 
entropy~\cite{Wald},
\begin{equation}
S_W=\frac{\mathcal A_k(r_H)\,F_R(R_H)}{4 G_N}\,,\quad
\Delta S_W=\frac{1}{4G_N}\left(2V_k r_H F_R(R_H)dr_H+V_k r_H^2 F_{RR}(R_H)d R_H\right)\,.
\end{equation}
The second expression above holds true when $R_H$ is an explicit function of $r_H$ only.
The condition on the entropy looks quite restrictive, but in a large class of explicit examples of $F(R)$-static black hole solutions it is well satisfied.
Thus, we can derive for a generic $F(R)$-gravitational model a
First Law of black hole thermodynamics in the form 
\begin{equation}
\Delta E_{K}:=T_{K}\Delta S_W\,,
\end{equation}
where $\Delta E_{K}$ is the variation of the Killing energy of the black hole itself. As a consequence,
one may define 
\begin{equation}
 E_{K}:=\frac{\mathcal V_k}{4\pi}\int\,\text{e}^{\alpha(r_H)}
\left(\frac{k\,F_R(R_H)}{2G_N}-\frac{R_HF_R(R_H)-F(R_H)}{4 G_N}r_H^2\right)d r_H\,.
\end{equation}
Here, an expression for the  BH-energy is proposed by deriving
the First Law from the equations of motion of $F(R)$-gravity, 
evaluating independently the entropy via Wald method and the 
Hawking temperature via quantum mechanical methods in curved space-times.

Let us consider some examples where only one integration constant $C$ appears in the SSS metric which may decsribes a black hole for some choices of the topology.
For the case $R=4\Lambda$ with $\Lambda=(R F_R(R)-F)/(2F_R(R))$ and 
Schwarzshild dS/AdS solution $\alpha(r)=0$, $B(r)=\left(k-C/r-\Lambda r^2/3\right)$, we get
\begin{equation}
T_K=\frac{4\pi r_H}{(1-\Lambda r_H^2)}\,,\quad
S_W=\frac{\mathcal A_k(r_H)F_R(R_H)}{4 G_N}\,, \quad
E_K=
\frac{\mathcal V_kF_R(R_H)}{8\pi G_N}r_H\left(k-\frac{\Lambda}{3}r_H^2\right)\,.
\end{equation}
Therefore, by using the fact that $B(r_H)=0$, one has
\begin{equation}
 E_K=\frac{\mathcal V_k F_R(R_H)}{8\pi G_N}\,C\,.
\end{equation}

For the model $F(R)=\gamma\sqrt{k(R+12\lambda)}$ with $\alpha(r)=0$ and $B(r)=\left(\frac{k}{2}-\frac{C}{r^2}+\lambda r^2\right)$, the BH Killing energy reads
$E_K\propto C$.

For the model $F(R)=\gamma(1/R-h^2/6)$ with $\exp\left[2\alpha\right]=r/r_0$, $r_0$ being a dimensional parameter, and $B(r)=\frac{4}{7}\left(k-\frac{7r}{6h}
+\frac{C}{r^{7/2}}\right)$, one obtains $E_K\propto C$.

For the class of Clifton-Barrow models $F(R)=R^{\delta+1}(\kappa^2)^\delta$, $\delta\neq 1$, the metric reads~\cite{CB},
\begin{equation}
ds^2=-\left(\frac{r}{r_0}\right)^{2a}\left(k-\frac{C}{r^b}\right)dt^2+\frac{\beta dr^2}{\left(k-\frac{C}{r^b}\right)}+r^2 d\Omega^2_k\,,
\end{equation}
where $a, b$ and $\beta$ are functions of $\delta$. Also in this case
the BH Killing energy results to be
$E_K\propto C$.
In all this examples, the Killing energy is proportional to the integration constant of the metric, giving to
it a physical meaning like in the Schwarzwschild case of GR. We point out that, when $\alpha(r)\neq 0$, if one uses the Hayward prescription 
such a result cannot be achieved, but with the Killing formalism there are some sorts of cancellations and we obtain this reasonable results. 

Let us consider the case of $R^2$-gravity, where two integration constants appear in the metric. The action reads:
\begin{equation}
I=\int_{\mathcal{M}} d^4x\sqrt{-g}\,R^2\,.
\end{equation}
Such a model is often considered in the inflationary scenario and admits the Schwarzshild dS/AdS solution,
\begin{equation}
 \alpha(r)=0\,,\quad B(r)=\left(k-\frac{C}{r}-\frac{\lambda r^2}{3}\right)\,,
\end{equation}
where $R=4\lambda$ and the cosmological constant $\lambda$ is a free parameter like $C$, due to the fact that it is not fixed by the gravitational Lagrangian. As a consequence, when we take the thermodynamical variation of 
the Killing energy of the black hole described by this solution, 
we must consider the variation with respect to $\lambda$ also and it is not possible to give an explicit expression for the energy. 
On the other hand, the Wald entropy of the black hole reads,
\begin{equation}
S_W=\frac{\mathcal A_k(r_H) R_H}{2}=2\mathcal A_k(r_H)\lambda\,,
\end{equation}
and vanishes for $\lambda=0$. The cosmological constant plays the role of the inverse of the Planck Mass of GR, since the action is scale invariant. 
In fact, $\lambda=1/L^2$ introduces a fundamental lenght scale into the theory and one may consider it like a fixed parameter. 
Only in this case the First Law leads to
\begin{equation}
E_K=\frac{\mathcal V_k}{\pi}\lambda\left[k r_H-\frac{\lambda r_H^3}{3}\right]=\frac{\mathcal V_k}{\pi}\lambda C\,.
\end{equation} 
We observe that the presence of the $R^2$-term modifies the energy of a Schwarzschild dS/AdS black hole when the cosmological constant is different to
zero (for example, in the model with Lagrangian $\mathcal L= (R-2\Lambda)/(16\pi G_N)+R^2$).

\section{Gauss-Bonnet modified gravity}

Let us consider now the following action,
\begin{equation}
I=\frac{1}{16\pi G_N}\int_\mathcal{M} d^4\sqrt{-g} F(R,G)\,,
\end{equation}
where $F(R,G)$ is a function of the Ricci scalar $R$ and the Gauss-Bonnet four dimensional topological invariant $G$. 
In this framework some static SSS black hole solutions are known, but in general it is not possible to derive the First Law from the field equations of the theory. However, given
a BH solution, it is still possible
to evaluate its Killing temperature, its Wald entropy and therefore its Killing energy.
For example, the model with $F(R, G)=R+\sqrt{G}$ admits the topological SSS solution,
\begin{equation}
ds^2=-B(r)dt^2+\frac{dr^2}{B(r)}+r^2d\,\left(\frac{d\rho^2}{1-k\rho^2}+\rho^2 d\phi^2\right)\,,\quad B(r)=-k+\frac{r}{C}\,,
\end{equation} 
$C$ being an integration constant, which describes a black hole in the spherical topological case with $k=1$. The Wald entropy reads
\begin{equation}
S_W=\frac{\mathcal A_1(r_H)}{4 G_N}\left[\mathcal F'_R+\mathcal F'_G\left(\frac{4k}{r^2}\right)\right]\Big\vert_H=
\frac{\pi r_H^2}{G_N}\left[1+\frac{C}{r_H}\right]\,,
\end{equation}
and we see that, since $\partial_r S_W|_{r_H}\neq \Delta S$, the First Law can not been derived from the equations of motion like in $F(R)$-gravity.
On the other side, by using the First Law with the Killing temperature we find
\begin{equation}
T_K=\frac{1}{4\pi C}\,,\quad E_K=\frac{C}{G_N}\,,
\end{equation}
and we see that even in this case the integration constant of the solution can be identified with the energy. We may conclude that in the vacuum case of $F(R)$ and $F(R,G)$-gravity
the Killing formalism leads to resonable definitions for the thermodynamics of the black holes.

\section{Non vacuum static spherically symmetric solution}

As a last example, we will consider the following model where a scalar field $\phi$ is non-minimally coupled with the electromagnetic potential ($F_{\mu\nu}$ is the 
electromagnetic stress-energy tensor):
\begin{equation}
I=\int_{\mathcal{M}} d^{4}x\sqrt{-g}\left[\frac{(R-2\Lambda)}{16\pi G_N}-\frac{1}{2}
\partial^\mu\phi\partial_\mu\phi
+V(\phi)
-\xi\mathrm{e}^{\sqrt{16\pi G_n}\lambda\phi}(F^{\mu\nu}F_{\mu\nu})\right]\,,\quad 
V(\phi)=V_0\mathrm{e}^{\gamma\sqrt{16\pi G_N}\phi}\label{V}\,.
\end{equation}
Here, $\Lambda\,, \lambda$, $\xi\,, \gamma$ and $V_0$ are fixed parameters of the theory. This model admits the following class of topological Lifshitz-like solutions,
\begin{equation}
ds^2=-\left(\frac{r}{r_0}\right)^{z}B(r)dt^2+\frac{d r^2}{B(r)}+r^2d\Omega_{k}^2\,,\label{dsds}
\end{equation}
where $z$ is a number and $r_0$ a dimensional parameter. The equations of motion constrain the field $\phi=\phi(r)$ as
\begin{equation} 
\phi(r)=\sqrt{\frac{2z}{16\pi G_N}}\log[r/ r_0]\,.
\end{equation}
The form of $B(r)$ results to be,
\begin{eqnarray}
\hspace{-1cm}B(r)=\frac{2k}{z+2}-\frac{C}{r^{1+\frac{z}{2}}}+\frac{\tilde V_0 r^2}{(2\gamma\sqrt{2z}+6+z)}
\left(\frac{r_0}{r}\right)^{\lambda\sqrt{2z}}
+\frac{8\xi\tilde Q^2 }{(2\lambda\sqrt{2z}+2-z)r^2}
\left(\frac{r}{r_0}\right)^{\gamma\sqrt{2z}}
-\frac{4\Lambda r^2}{6+z}\,,\label{BB}
\end{eqnarray}
where $C$ is a free integration constant of the solution, $\tilde V_0=16\pi G_N V_0$, 
$\tilde Q^2=G_N Q^2$,
$Q$ being the charge of the electromagnetic potential, and the parameters of the model must be related to each other in order to satisfy the
Klein-Gordon equation of the scalar field.
For $\phi(r)=0$, namely $z=0$, $\xi=1/4$ and $V_0=0$, we recover the Reissner-Norstrom solution with cosmological constant. 

The model under investigation has second order field equations like in GR and when solution (\ref{dsds}, \ref{BB}) describes a black hole its mass is well defined as
\begin{equation}
E=\frac{\mathcal V_{k}}{8\pi G_N}C \,,
\end{equation}
while the entropy satisfies the Area Law (\ref{AreaLaw}).
Now from the first component of the field equations evaluated on the BH horizon we derive
\begin{equation*}
\frac{1}{4\pi}B'(r_H)\Delta S=\Delta E+pdV\,,\quad p=(p_\phi+p_\text{EM})_\text{radial}\,,
\end{equation*}
where the working term collects the contributions from the radial pressures of the scalar and electromagnetic fields.
Thus, the First Law holds true by making use of the Hayward temperature in (\ref{TK}). It looks that when we consider a static but non vacuum solution the Hayward formalis is 
more adapt to describe the thermodynamics of the black holes.

\end{document}